\documentclass[12pt]{article}

\usepackage{url}
  \usepackage[dvipsnames]{xcolor}
  \usepackage{adjustbox, setspace}
  \usepackage{caption,subcaption,wrapfig}
\usepackage[normalem]{ulem} 
 \useunder{\uline}{\ul}{} 
 \usepackage{amsmath,multirow, bm}
 
 \usepackage{tikz}
 \usepackage{pgf}
\usetikzlibrary{shapes.geometric, arrows, positioning}
\usetikzlibrary{decorations.pathreplacing}
 \usepackage{natbib}

\newtheorem{theorem}{Theorem}
\newtheorem{corollary}[theorem]{Corollary}

\newtheorem{proposition}[theorem]{Proposition}
\newtheorem{definition}[theorem]{Definition}

 \newcommand\ben{\begin{enumerate}}
 \newcommand\een{\end{enumerate}}
  \newcommand\beq{\begin{eqnarray}}
 \newcommand\eeq{\end{eqnarray}}
   \newcommand\beqa{\begin{eqnarray*}}
 \newcommand\eeqa{\end{eqnarray*}}
   \newcommand\bal{\begin{aligned}}
 \newcommand\eal{\end{aligned}}
  \newcommand\bi{\begin{itemize}}
 \newcommand\ei{\end{itemize}}

\newcommand{\0}{\bm{\0}}

\begin{document}

\title{Surrogate-guided sampling designs for classification of rare outcomes from electronic medical records data}
\date{}
\author{W. KATHERINE TAN$^\ast$, PATRICK J. HEAGERTY
\\[2pt]
\textit{Department of Biostatistics}\\
 \textit{University of Washington}\\
 \textit{Seattle Washington USA}
 \\[2pt]
 }

\markboth%
{} 
{Surrogate-guided sampling}

\maketitle


\begin{abstract}
 {\noindent Scalable and accurate identification of specific clinical outcomes has been enabled by  machine-learning applied to electronic medical record (EMR) systems. The development of classification models requires the collection of a complete labeled data set, where true clinical outcomes are obtained by human expert manual review. For example, the development of natural language processing algorithms requires the abstraction of clinical text data to obtain outcome information necessary for training models. However, if the outcome is rare then simple random sampling results in very few cases and insufficient information to develop accurate classifiers.  Since large scale detailed abstraction is often expensive, time-consuming, and not feasible, more efficient strategies are needed. Under such resource constrained settings, we propose a class of enrichment sampling designs, where selection for abstraction is stratified by auxiliary variables related to the true outcome of interest. Stratified sampling on highly specific variables results in targeted samples that are more enriched with cases, which we show translates to increased model discrimination and better statistical learning performance. We provide mathematical details, and simulation evidence that links sampling designs to their resulting prediction model performance.  We discuss the impact of our proposed sampling on both model training and validation. Finally, we illustrate the proposed designs for outcome label collection and subsequent machine-learning, using radiology report text data from the Lumbar Imaging with Reporting of Epidemiology (LIRE) study.}
 {Electronic medical records, Machine learning, Observational Studies, Sampling design.}
\end{abstract}

\section{Introduction}
\label{sec:sgs_intro}
\noindent Linked electronic medical record (EMR) systems provide a massive reservoir of information that can help researchers understand and treat both common and rare medical conditions. Specifically, EMR data includes both \textit{structured} data, such as lab values and diagnostic codes, and \textit{unstructured} data in the form of free-text medical notes and images. In order to extract research ready variables, ultimately both structured data and carefully processed unstructured data are necessary, but extracting specific findings from unstructured data is often expensive and time-consuming. The traditional manual abstraction approach requires highly trained clinicians or technicians transcribing medical notes into coded variables, and is not scalable to massive EMR cohorts. As scalable alternatives, machine-learning methods have been developed, for example natural language processing (NLP) methods for medical text data (\cite{chapman2001comparison, carroll2012portability}), and deep learning strategies for medical images (\cite{esteva2017dermatologist}). Yet, any algorithm development relies on a base of training and validation data, and the purpose of this manuscript is to outline efficient study designs that can facilitate cost-effective data collection for the development of new prediction tools.\\

\noindent In order to both develop (i.e. train) and validate (i.e. test) data-driven machine-learning algorithms, a sample containing both feature (predictor) and outcome information (``labeled data") is required. In typical EMR settings, labeled data is not readily available, therefore a small subset of the underlying cohort needs to be selected for outcome abstraction. A challenge for machine-learning of clinical outcomes is class distributional ``imbalance" where cases (outcome=1) are disproportionately less frequent than controls (outcome=0). For classification tasks, the training sample outcome class distribution has been demonstrated to affect classification accuracy, both empirically (\cite{weiss2001effect, batista2004study, wei2013role}) and theoretically (\cite{xue2015does}). To address  class imbalance, one approach involves re-sampling the training sample to eliminate controls (under-sampling) or replicating cases (over-sampling), in order to re-balance the effective outcome class distribution in training data, and hopefully to improve ultimate model prediction accuracy (\cite{chawla2002smote, he2009learning}). However, such \textit{analysis}-based re-sampling procedures assume that an initial labeled data sample is already available, and these strategies disregard the potential cost associated with labeled data collection (\cite{weiss2001effect}).\\

\noindent When data collection resources are scarce, targeted sampling methods in epidemiology have offered highly efficient research designs. In contrast to \textit{analysis}-based re-sampling procedures, epidemiologic sampling methods are defined at the \textit{design} stage of studies prior to data collection. A well-known example is the case-control design (\cite{prentice1979logistic}), where expensive data ascertainment is based on strata defined by values of a cheaper auxiliary variable and may be viewed as special cases of the general two-phase sampling design (\cite{neyman1934two, chatterjee2003pseudoscore}). In the context of effect estimation, targeted sampling through two-phase designs has been shown to provide efficiency over simple random sampling (\cite{zhao2009likelihood, mcisaac2014response}), especially when using sampling variables that are highly correlated and informative for the outcome (\cite{zhao2012design}). However, the effect of selectively sampled training data on ultimate machine-learning prediction accuracy has not been thoroughly investigated.\\

\noindent For clinical outcome identification using EMR data, an imperfect alternative to abstracted outcomes may be based on summaries of related structured data elements, such as International Classification of Disease (ICD) codes and simple keyword searches queried within pre-specified time frames. Such ``surrogates" or ``correlates" of actual clinical outcomes have been used in place of true clinical outcomes in machine-learning modeling tasks to reduce the dimensionality of EMR-generated features (\cite{yu2016surrogate, gronsbell2019automated}), or directly as ``noisy" imputed outcome labels for classifier development (\cite{agarwal2016learning}). Alternatively, surrogates could help guide selection of subjects for labeled data abstraction, for example selecting subjects using non-negated keywords and ICD codes to assemble labeled data (\cite{pakhomov2005prospective}). Yet, there remains little discussion of corresponding statistical rationale, and such heuristic decisions based on purposeful biased sampling may not create generalizable predictions, or valid summaries of accuracy.\\

\noindent This paper is motivated by the need for a formal statistical framework to guide sampling of subjects for labeled data abstraction, towards accurate and scalable machine-learning classification of clinical outcomes. We specifically focus on the rare outcome scenario, where model accuracy is often rate-limited by the number of outcome cases. As with conventional intuition, our proposed strategy targets case-enrichment of rare outcomes for selection of training data. The key contribution of our work is the formalization of heuristic sampling methodologies drawn from the fields of machine-learning and epidemiology, therefore filling a critical gap in EMR research methods.

\section{Methods}
\label{sec:sgs_methods}
\subsection{Statistical motivation and proposed design}
\label{sgs_sec:methods_1}

\noindent For subject $i$ denote $\tilde{X}_i \in \mathcal{R}^p$ as the feature vector consisting of features $X_{ij}$, $j=1,\hdots,p$ and $Y_i \in \{0,1\}$ as the binary outcome. The general classification problem is to find a function $h(.)$ that maps from the features to outcomes, for example logistic regression with and without regularization

\beq\bal 
\hat{\beta_0}, \hat{\tilde{\beta}}_X &= 
\underset{\beta_0, \tilde{\beta_X}}{min} \{-\sum\limits_{i=1}^{n} Y_{i}(\beta_0 + \sum\limits_{j=1}^{p} \beta_{X_j} X_{ij})
+\log(1 + \exp(\beta_0 + \sum\limits_{j=1}^{p} \beta_{X_j} X_{ij}))
+ \lambda \sum\limits_{j=1}^{p} ||\beta_{X_j}||_L \}.
\label{sgs_eq:regularized_logistic_regression}
\eal\eeq

\noindent where in \eqref{sgs_eq:regularized_logistic_regression}, $L=1$ refers to Lasso regression (\cite{tibshirani1996regression}), $L=2$ refers to Ridge regression (\cite{le1992ridge}), and $\lambda=0$ is equivalent to logistic regression without regularization. For a concrete example of application of classification models such as \eqref{sgs_eq:regularized_logistic_regression}, consider the task of classifying radiology reports for subject vertebral fracture status. For this NLP motivated task, features $\tilde{X}_i$ may be derived using bag-of-words (BOW) representations (for subject $i$, the BOW feature vector $\tilde{X}_{i}$ has binary elements $X_{ij} = I(t_j \in \text{report}_i)$ with unique terms $t_j$ obtained by concatenating all reports), while outcomes $Y_i$ must be obtained through abstraction (clinician-defined indicator of vertebral fracture). We consider sampling designs to select records for outcome abstraction so that both $\tilde{X}_i$ and $Y_i$ are available for machine-learning training and validation.\\

\noindent The typical assumption for sampling is that the sample is obtained through simple random sampling (SRS) from a specified target population. An alternative to SRS is a targeted sample enriched specifically to improve machine-learning performance. For example, often times there exists other structured data elements in EMR databases that are related to $Y_i$. For instance, clinician-identified vertebral fracture may be related to keywords representing fracture in report text, or ICD codes recorded during the same subject visit. Denote summaries of such related structured data elements as $Z_i$, which we define as ``surrogates" for the true outcome. Note that $Z_i$ may be a subset of features $\tilde{X}_{i}$ in \eqref{sgs_eq:regularized_logistic_regression}.

\begin{definition} Surrogate-guided sampling (SGS) design class.\\
Denote the surrogate-guided sampling (SGS) design class as the set of stratified sampling procedures based only on values of a binary enrichment surrogate $Z \in \{0,1\}$.  Such designs would select an individual $i$ for sampling with probability $\pi(Z_i)$ where typically $\pi(Z_i=1) > \pi(Z_i=0)$ when $Z$ is positively correlated with $Y$.
\label{sgs_def:sgs}
\end{definition}

\noindent The surrogate-guided sampling (SGS) design class (Definition \ref{sgs_def:sgs}) describes the class of stratified sampling designs based on values of an enrichment surrogate, and is a special case of two-phase sampling. In SGS designs, all subjects in the cohort are divided into two strata based on surrogate values: surrogate positives with $Z_i=1$, and surrogate negatives with $Z_i=0$. Then, subjects are selected into the sample based on surrogate values, and only selected subjects have true $Y_i$ abstracted for. The intended benefit of SGS designs is that, for the same abstraction cost, resulting samples have higher expected outcome prevalences compared to using SRS. For illustration, consider an outcome prevalence of 10\%, and assume that in the EMR, there exists a surrogate with 40\% sensitivity and 95\% specificity for the outcome of interest. The outcome prevalence in the surrogate positive and negative strata are expected to be approximately 47\% and 6.6\% respectively, corresponding to the surrogate positive and negative predictive values. Then, for an abstraction budget allowing collection of 500 labels, using an SGS with 1:1 ratio of surrogate positives to negatives design (i.e. ``balanced" design) yields 134 cases in expectation. In contrast, an SRS design would have required abstraction of 1340 subjects to yield 134 cases, constituting an abstraction burden of more than 2.5 times. Note that cases identified using SGS designs are true cases collected from the cohort, and not replicates or synthetic data as resulting from using analysis-based re-balancing methods.

\subsection{Effect of training sample composition on prediction accuracy}
\label{sgs_sec:methods_2}

\noindent To demonstrate that the sampling design choice does affect finite-sample learning performance, we provide a mathematical representation of how training sample composition impacts prediction accuracy. For tractability we focus on a commonly used evaluation metric, the Area Under the Receiver Operating Characteristic (ROC) Curve (AUC). Model validation AUC can be interpreted as how well resulting continuous predictions discriminate between randomly selected pairs of case and control subjects in yet unseen data. Other performance metrics such as binary accuracy correspond to the sum of error values for a particular point on the ROC curve. If we consider continuous model predictions as a ``test'' for true outcome statuses, then assuming that ``test" conditioned on outcomes are normally distributed (``bi-normality"), \cite{pepe2003statistical} has shown the AUC to be

\beq\bal 
AUC = \Phi(\sqrt{R_{AUC}}) = \Phi \left( \sqrt{\dfrac{(\mu_1 - \mu_0)^2}{\sigma_1^2 + \sigma_0^2}} \right).
\label{sgs_eq:binormalAUC}
\eal\eeq 

\noindent In \eqref{sgs_eq:binormalAUC}, $\mu_y$ and $\sigma_y^2$ are the means and variances of the ``test'' among the cases ($y=1$) and controls ($y=0$). The bi-normal AUC formula \eqref{sgs_eq:binormalAUC} was developed in \cite{pepe2003statistical} for diagnostic testing applications, but may be generalized to the classification modeling setting. For classification model development, continuous model predictions are estimated using a training sample, and generalizable performance usually evaluated on a separate validation sample. Denote $\mathbf{D}^{S}(n)$ as the training sample collected using sampling design $S$ and having sample size $n$, and assume that the validation sample is a large sample obtained through SRS from a population $\mathcal{D}$. Then, the validation AUC for model developed with $\mathbf{D}^{S}(n)$ may be represented using an indexing as shown in Definition \ref{sgs_def:auc_index}.

\setcounter{theorem}{1}
\begin{definition}
\label{sgs_def:auc_index} $AUC(Y|\mathbf{D}^{S}(n))$.\\
Let $AUC(Y|\mathbf{D}^{S}(n))$ denote the validation AUC of a classification model for outcome $Y$ developed using sample $\mathbf{D}^{S}(n)$ defined with sampling design $S$ and sample size $n$.
\end{definition}

\noindent Using the indexing as in Definition \ref{sgs_def:auc_index} to represent validation AUC in terms of training sample composition, Theorem \ref{sgs_thm:1} shows that $AUC(Y|\mathbf{D}^{S}(n))$ is inversely proportional to the estimation variance and the data signal-to-noise ratio (details in Supplementary Material A). Therefore, assuming use of the same modeling procedure, using a design with higher statistical information as measured by lower estimation variance results in higher $AUC(Y|\mathbf{D}^{S}(n))$. To our knowledge, the results in Theorem \ref{sgs_thm:1} are the first to directly present an indexing of validation AUC in terms of training sample composition. We may use the results in Theorem \ref{sgs_thm:1} to explain the effect of outcome class imbalance on classifier discrimination. For example when modeling using logistic regression, samples with rare outcomes tend to result in more highly variable coefficient estimates compared to that of more prevalent outcomes (\cite{king2001logistic}) -- such increased estimation variance is related to lower discrimination. Note that model prediction bi-normality may be obtained for features $\mathbf{X}|Y=y$ that are bi-normal, or monotone transformations of normal distributions (\cite{pepe2003statistical}). In addition, the results in Theorem \ref{sgs_thm:1} may be generalized beyond logistic regression to include regularization, as long as the estimation bias and variance of resulting coefficients can be well characterized.

\setcounter{theorem}{0}
\begin{theorem}
\label{sgs_thm:1}
\noindent Assume that in population $\mathcal{D}$, for $y \in \{0,1\}$, $\mathbf{X}|Y = y$ has mean $\mu_{x|y}$ where $\mu_{x|y=0} = 0$, and covariance $\mathbf{\Sigma}_{x|y} = \mathbf{\Sigma}_{x|y=1} = \mathbf{\Sigma}_{x|y=0}$. Let the estimated model predictions $\hat{\eta} = \mathbf{X}\hat{\beta}$ be bi-normally distributed such that $\hat{\eta} \sim N(\mu_y, \mathbf{\Sigma}_y)$, and model coefficients $\hat{\beta}$ are estimated by logistic regression using training sample $\mathbf{D}^S(n)$. Then, 

\beq\bal 
AUC(Y|\mathbf{D}^S(n)) \propto \dfrac{1}{trace(\mathbf{\Sigma}_{x|y} \bm{V}{(\hat{\beta}^S(n))} + \mu_{x|y=1}^T \bm{V}{(\hat{\beta}^S(n))} \mu_{x|y=1}},
\label{sgs_eq:auc_simplification}
\eal\eeq 

\noindent where $\bm{V}{(\hat{\beta}^S(n))} = (\mathbf{X}^T \mathbf{W} \mathbf{X})^{-1}$ is the approximate covariance matrix associated with estimating $\hat{\beta}$ using $\mathbf{D}^S(n)$, and $\mu_{x|y=1}$ and $\mathbf{\Sigma}_{x|y}$ are parameters describing the data signal-to-noise ratio.
\end{theorem}
\subsection{Factors affecting SGS design characteristics}
\label{sgs_sec:methods_3}

\noindent SGS design characteristics, specifically surrogate operating characteristics and strata proportions, both affect sample information. Motivated by empirical results in machine-learning, we use sample outcome prevalence as a simple measure of information. We first discuss optimal strata proportions towards maximizing sample outcome prevalence, and then propose an alternative framework to demonstrate key drivers for high information SGS designs.

\subsubsection{Optimal strata proportions for given desired sample outcome prevalence}

\noindent Let $S=1$ denote sampling into sample $\mathbf{D}^S(n)$. To obtain a desired sample outcome prevalence in $\mathbf{D}^S(n)$, the optimal proportion of surrogate positives in the sample is

\beq\bal 
R_{opt} &= \dfrac{P(Y=1|S=1) + NPV_Z - 1}{PPV_Z + NPV_Z -1}, R_{opt} \in [0,1]
\label{sgs_eq:optimal_r}
\eal\eeq 

\noindent derived using a simple application of Bayes rule (details in Supplementary Material B), where $P(Y=1|S=1)$ is the desired prevalence in $\mathbf{D}^S(n)$ and $PPV_Z$, $NPV_Z$ are the positive and negative predictive values of surrogate $Z$ respectively. Consider a surrogate with sensitivity of 40\% and specificity of 95\% for the true outcome and we desire a sample prevalence of 50\%. Then when the natural outcome prevalence in the cohort is 20\% $R_{opt}$ is 0.69, indicating that ideally about two-thirds of $\mathbf{D}^S(n)$ should be surrogate positives. However, such a desired sample prevalence cannot be achieved for natural outcome prevalences of 11\% or less, since $R_{opt}$ is required to be between 0 and 1. For such rare outcome scenarios, we next describe a alternative framework to determine high information SGS designs in terms of both surrogate operating characteristics and strata proportions.

\subsubsection{Properties of design $O_{ratio}$ and the impact of surrogate specificity}

\noindent For scenarios where outcome prevalences are less than 50\%, consider the transformation of sample outcome prevalence into sample case/control odds where higher odds indicate higher prevalence. To denote the sample case enrichment comparing SGS to SRS, we propose using the case/control odds ratio, a metric we denote as $O_{ratio}$ and mathematically define in Definition \ref{sgs_def:oratio}.

\setcounter{theorem}{2}
\begin{definition} \noindent $O_{ratio}$.\\
\label{sgs_def:oratio}
Let $O_{ratio}$ denote the expected case/control odds ratio comparing SGS to SRS, where 
$O_{ratio} = \frac{E^{\mathbf{D}^{SGS}(n)}[Y|S=1]}{1-E^{\mathbf{D}^{SGS}(n)}[Y|S=1]} /\frac{E^{\mathbf{D}^{SRS}(n)}[Y|S=1]}{1-E^{\mathbf{D}^{SRS}(n)}[Y|S=1]} = \dfrac{\text{Odds}(\text{cases}|SGS)}{\text{Odds}(\text{cases}|SRS)}$. 
\end{definition}

\noindent The denominator of $O_{ratio}$ is the expected odds of cases for samples collected with SRS, and is less than $1$ for outcome with prevalences less than 50\%. The numerator is the expected odds of cases for samples collected with SGS designs. Therefore, $O_{ratio}$ is a single estimate of design effect on sample outcome prevalence, and can be interpreted as the expected increase in cases comparing SGS to SRS, with higher values indicating that SGS provides more case enrichment, and $O_{ratio} > 1$ indicating improvement using SGS relative to SRS. An interesting property of $O_{ratio}$ is the connection to Likelihood Ratios (LRs) of the enrichment surrogate. Of note, LRs of a diagnostic test can be interpreted as slopes of Receiver Operating Characteristics (ROC) curves, are related to positive and negative predictive values (PPV \& NPV), but are invariant to outcome prevalence (\cite{choi1998slopes}). Therefore, by framing enrichment surrogates $Z$ as ``prior tests" of outcome $Y$, we may gain insight into what types of variables are the best surrogates for sampling.

\setcounter{theorem}{0}
\begin{proposition} Properties of $O_{ratio}$.\\
Let an SGS design of sample size $n$ be defined with surrogate $Z$ and sampling ratio $R=P(Z=1|S=1)$, where $R=0.50$ corresponds to a ``balanced" SGS 1:1 design. Let $Z$ has operating characteristics: $Z_{sens} := P(Z=1|Y=1)$, $Z_{spec} = P(Z=0|Y=0)$. Then, if the outcome is rare ($P(Y=1) \approx 0$), then

\beq\bal
O_{ratio}(R,Z) &\approx (R) (LR+) + (1-R) (LR-). \label{sgs_eq:oratio_linear}
\eal\eeq
\label{sgs_prop:1}
\end{proposition}

\setcounter{theorem}{0}
\begin{corollary}
\noindent For a given $Z$, $O_{ratio} \propto R$. Over the set of possible $Z$, $O_{ratio} \propto Z_{sens}$ and $O_{ratio} \propto \dfrac{1}{1-Z_{spec}}$.
\label{sgs_cor:1}
\end{corollary}

\begin{figure}[!htbp]
\centering
\caption{$O_{ratio}$ values for surrogates of different marginal sensitivity and specificity, based on a fixed $R=0.50$ and an outcome with prevalence of $10\%$.}
\includegraphics[width=4.5in,height=2.5in]{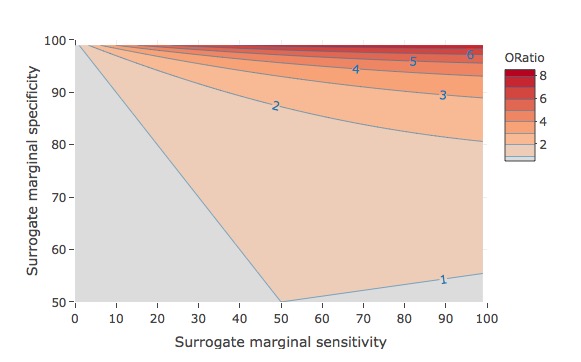} 
\label{sgs_fig:oratio_surface}
\end{figure}

\noindent Details of Proposition \ref{sgs_prop:1} and Corollary \ref{sgs_cor:1} are in Supplementary Material B. For a given surrogate, higher values of $O_{ratio}$ can be achieved by over-representing surrogate positives. Over the range of possible surrogates, a small change in surrogate specificity can have a much higher impact on $O_{ratio}$ compared to the same change in surrogate sensitivity. Figure \ref{sgs_fig:oratio_surface} demonstrates the impact of surrogate operating characteristics on $O_{ratio}$ for a fixed sampling ratio of $R=0.50$ (i.e. SGS 1:1 ratio of cases to control, a ``balanced" design) and prevalence of 10\%. An SGS design with $O_{ratio} > 2$, may be obtained by using a surrogate with (sensitivity, specificity) of $(30\%, 90\%)$ or $(90\%, 80\%)$, where in order to maintain the same sample case-control odds a small decrease of surrogate specificity required a much larger increase in surrogate sensitivity. In fact, very high information designs, such as $O_{ratio} > 3$, can only be achieved when surrogate specificity is at least 90\%.

\subsection{Design impact on model training and model validation}
\label{sgs_sec:methods_4}

\noindent To improve the information of samples selected for machine-learning, SGS designs intentionally over-represent surrogate positives. The impact of such sample selection bias on machine-learning was first formalized in \cite{zadrozny2004learning}, and can be formulated as a missing data problem (\cite{little2014statistical}). Specifically, since SGS sampling only depends on surrogate values $Z$ that is available for all subjects in $\mathcal{D}$, sampling is independent of outcome labels conditional on surrogate values ($S \perp Y | Z$) -- we now describe the impact of such a Missing At Random (MAR) assumption on both model training and model validation. 

\subsubsection{Design impact on model training}

\noindent To characterize design impact on model training, we consider the impact of using training sample $\mathbf{D}^S(n)$ for asymptotically unbiased estimation of the true model. Specifically, when sampling design $S$ induces an MAR assumption, \cite{zadrozny2004learning} had suggested two conditions:  first, estimated models need to depend asymptotically only on conditional outcome distributions $f(y|x,z)$ and not feature distributions $f(x,z)$ and second, sampling variables need to be included in model estimation. The first condition is generally met by regression models such as regularized logistic regression and generalized additive models, but not for models such as decision trees and Bayesian models (\cite{zadrozny2004learning}). The second condition can be achieved by requiring the surrogate variable to be included in fitted models, such as assigning a penalty of zero to the surrogate variable for L1 regularized regression.

\subsubsection{Design impact on model validation}

\noindent Often times in practice it may be operationally advantageous to collect outcome labels using a single sampling design, and then split data into separate training and validation samples. We characterize the design impact on model validation, specifically in using sample $\mathbf{D}^S(n)$ to assess model generalizable prediction accuracy metrics such as sensitivity, specificity, and AUC. In general, unless the validation sample is drawn randomly from the cohort (i.e. SRS), empirically estimated accuracy metrics are typically biased for the true values. Since SGS designs are MAR, empirical estimates may be adjusted using inverse sampling probabilities (\cite{horvitz1952generalization}). In the validation sample, for subject $i$ let $\hat{p}_i$ be the model-predicted probability and $\pi_i$ be the sampling probability, then Inverse Probability Weighted (IPW) estimator for AUC is 

\beq\bal
AUC_{IPW} &= \dfrac{\sum\limits_{i=1}^{n} \sum\limits_{j=1}^{n} \pi_i^{-1} \pi_j^{-1} I(\hat{p}_{i} > \hat{p}_{j}) I(Y_{i} > Y_{j})}{\sum\limits_{i=1}^{n} \sum\limits_{j=1}^{n} \pi_i^{-1} \pi_j^{-1} I(Y_{i} > Y_{j})}.
\label{sgs_eq:ipw_auc}
\eal\eeq  

\noindent In \eqref{sgs_eq:ipw_auc}, even though $\pi_i : \pi(Z_i)$ is known by construction for the SGS design, estimation from observed data may yield more efficient estimators.
 
\subsection{Extensions beyond binary surrogates}
\label{sgs_sec:methods_5}

\noindent The proposed surrogate-guided sampling design may be extended to scenarios beyond binary surrogates. One practical scenario multiple potential surrogate candidates, for example how to combine counts of relevant ICD codes and keywords for sampling -- we suggest three possible strategies. First, surrogate candidates may be assessed for specificity and the surrogate with highest specificity and sufficient surrogate positive sample size selected. Second, candidates may be combined using an ``AND" logic to create a highly specific composite surrogate. Third, cross-classifications may be created from candidate surrogates and sampling from each stratum could proceed using strategies from the two-phase literature, such as balanced sampling (\cite{breslow1999design}) having equal sample strata proportions, or optimizing sample strata proportions by minimizing the variance for the model used in analyses (\cite{mcisaac2014response}). Another practical scenario is extension to complex surrogates such as quantitative and multi-categorical variables, for example biomarker tests and raw counts of relevant ICD codes. For such situations, one strategy is to use unsupervised learning methods to summarize and combine surrogates. For example, sampling could be based on principle components with largest variations in order to obtain high information designs. Note that when using complex surrogates, sampling probabilities may not be known by design: using doubly robust estimators for model validation may relax requirements for correctly specifying sampling probabilities (\cite{wang2009causal}).

\section{Simulations}
\label{sec:sgs_sims}
\noindent To illustrate the benefit of using SGS designs for statistical machine-learning model training and validation, we conducted simulations motivated by a real-world data set of radiology text reports from the Lumbar Imaging with Reporting of Epidemiology (LIRE) study (\cite{jarvik2015lumbar} - for additional information see Section \ref{sec:sgs_data}) - where we generated features following a long-tail distribution that is characteristic of text data and conducted modeling using regularized regression due to high-dimensional assumptions. We demonstrated the impact of various sampling designs on model training and model validation.

\subsection{Design impact on model training}

\subsubsection{Set-up}

\noindent For cohorts of size $N=100,000$, we generated conditional outcomes as independent Bernoulli random variables, having prevalence of either 5\% or 10\%. As most EMR datasets contain features of high dimensionality, we set the number of features to be $p=250$, of which only $30$ had non-zero coefficients. Specifically, the conditional outcome was generated as $Y_i | \left( Z1_i, Z2_i , \tilde{X}^T_i\right) \sim \text{Bernoulli} (P(Y_i=1))$, where $logit(E[Y_i|Z1,Z2,\tilde{X}]) = \beta_0 + \beta_{Z1} Z1_{i} + \beta_{Z2} Z2_{i} + \sum\limits_{j=1}^{p} \beta_j X_{ij}$, with $\beta_j = (-0.75, -0.5, 0.25, \hdots, -0.5, 0.25)$ for the first $20$ most frequent features, $\beta_j = 1$ for the $10$ features with frequencies closest to the outcome prevalence, and $\beta_j = 0$ for the remaining $220$ features. Here, we used a simplifying assumption that the most predictive text-based features tend to occur as often as the outcome prevalence, frequent features are weakly predictive, but most features are irrelevant for predicting the outcome.\\

\noindent Binary features were generated as independent Bernoulli random variables, with marginal feature frequencies following an exponential distribution simulating a long upper tail distribution, where the most common features are present in almost all reports but the majority of features have very low frequencies (\cite{sichel1975distribution}). Specifically, features were generated as $\tilde{X}_{j} \sim \text{Bernoulli}\left(p_{\tilde{x}_j}\right)$, where $p_{\tilde{x}_j}$ simulated following an exponential distribution with mean = $\frac{1}{6}$ comparable to observed distributions in the LIRE dataset. For the binary enrichment surrogates, surrogate $Z1$ had a sensitivity of 40\% and a specificity of 95\%, defined to have comparable operating characteristics with the real-world surrogate for the LIRE data set, while surrogate $Z2$ had a sensitivity of 67\%  and a specificity of 66\%, and may be viewed as a ``weaker'' surrogate for sampling. Note that both surrogates have the same discrimination for the outcome (AUC = 0.67) as computed according to the trapezoidal rule.\\

\noindent We compared the following methods: SRS which we consider to be the ``baseline'', SGS, as well as random over-sampling (ROS) which is a commonly used analysis-based re-sampling procedure. For each simulated cohort, we set aside a large validation sample with sample size $n_{val} = 10000$ using SRS. From the remaining subjects, we simulated ``abstraction samples" varying across a grid of sample sizes, and sampling methods of SRS, ROS, SGS 1:1 (equal case to control ratio) or SGS 3:1 (cases are 3x controls), where SGS may be based on surrogates $Z1$ or $Z2$. For the SRS and SGS sampling designs, the abstraction sample size is exactly the training sample size. The ROS procedure replicates cases from an SRS sample of size $n$ until the number of cases and controls are equal. Therefore, even though both SRS and ROS have the same ``abstraction sample size", ROS results in a higher training sample size due to case replication. We used abstraction sample size, rather than training sample size, as the unit of cost measurement.\\

\noindent For each iteration we fit either Lasso or Ridge classification models, but coefficients for the surrogate used for sampling were assigned a zero penalty, which is a modification to the usual likelihood so that the surrogate is always included in the resulting model. Regularization parameters were selected based on values that maximized AUC using ten-fold cross-validation on training samples. Then, we apply resulting model estimates to the validation sample, calculating the empirical validation AUC using the Wilcoxon-Mann-Whitney formula. Over all $B=1000$ iterations, we calculated average validation AUCs and illustrated results in the form of learning curves. Briefly, a learning curve is a type of plot in machine-learning to show the change in model prediction accuracy (here: discrimination) when cost (here: abstraction sample size) increases. In these experiments, since we compared prediction accuracy across different sampling designs conditioned on the same models and data generating mechanism, the difference in model performance is due to differences in the sampling design that gave rise to resulting samples.

\subsubsection{Results}

\begin{figure}[!htbp]
\centering
\caption{Logistic Lasso Regression learning curves (outcome prevalence = 5\%) comparing simple random sampling (SRS), random over-sampling (ROS), and surrogate-guided sampling (SGS) with 1:1 or 3:1 ratio of surrogate positives to negatives. Surrogate $Z1$ had sensitivity = 40\% and specificity = 95\%, while surrogate $Z2$ had sensitivity = 67\% and a specificity = 66\%.}
\includegraphics[width=7in]{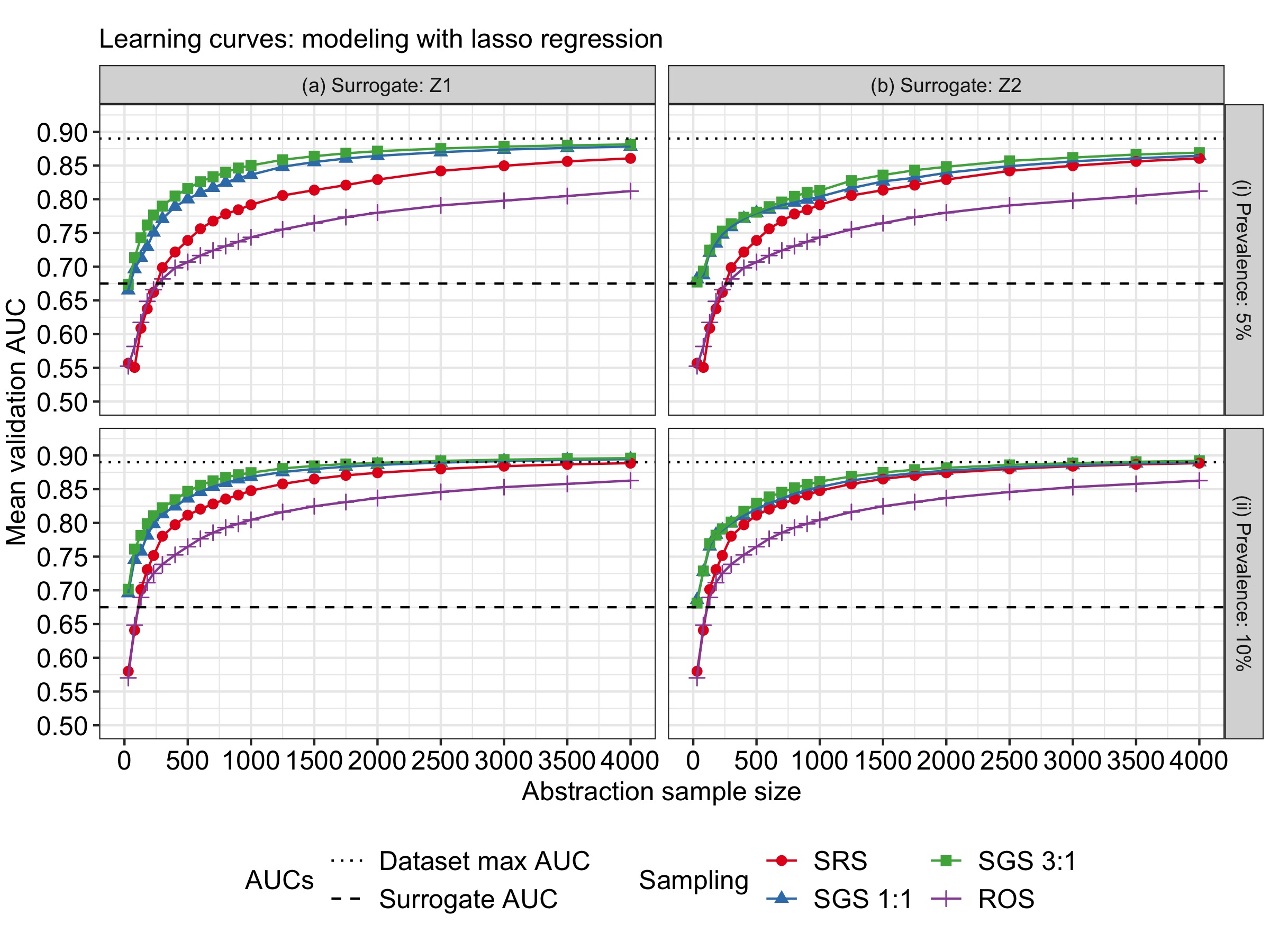}
\label{sgs_fig:cohort_lasso}
\end{figure}

\noindent Figure \ref{sgs_fig:cohort_lasso} illustrates simulation results when modeling with logistic lasso regression. First, consider the cohort with 5\% outcome prevalence and SGS sampling using surrogate Z1 (Figure \ref{sgs_fig:cohort_lasso}(a)(i)), where in order to achieve a validation AUC of 0.85 (94\% of the maximum AUC of 0.90), using SRS required an abstraction sample size of n=3000, while using SGS 1:1 required n=1500 (50\% of SRS cost) and SGS 3:1 required n=1000 (33\% of SRS cost). The impact of ROS on learning is inconsistent, where such case replication sometimes resulted in worse generalizable discrimination compared to no replication (SRS). Similar patterns were observed for SGS sampling with surrogate Z2 (Figure \ref{sgs_fig:cohort_lasso}(b)(i)). Even though surrogates Z1 and Z2 had the same discrimination for the outcome (AUC = 0.675), Z2 had lower specificity and was a weaker variable for stratified sampling purposes. To achieve a validation AUC of 0.85, using SGS 1:1 and SGS 3:1 allocations required n=2500 (83\% of SRS cost) and n=2000 (67\% of SRS cost) respectively. Similar results were observed for the 10\% outcome prevalence cohort (Figures \ref{sgs_fig:cohort_lasso}(a)(ii) and \ref{sgs_fig:cohort_lasso}(b)(ii)), but SGS design benefit over SRS was less pronounced due to a less rare outcome.\\

\begin{figure}[!htbp]
\centering
\caption{Logistic Ridge Regression learning curves (outcome prevalence = 5\%) comparing simple random sampling (SRS), random over-sampling (ROS), and surrogate-guided sampling (SGS) with 1:1 or 3:1 ratio of surrogate positives to negatives. Surrogate $Z1$ had sensitivity = 40\%  and specificity = 95\%, while surrogate $Z2$ had sensitivity = 67\%  and a specificity = 66\%.}
\includegraphics[width=7in]{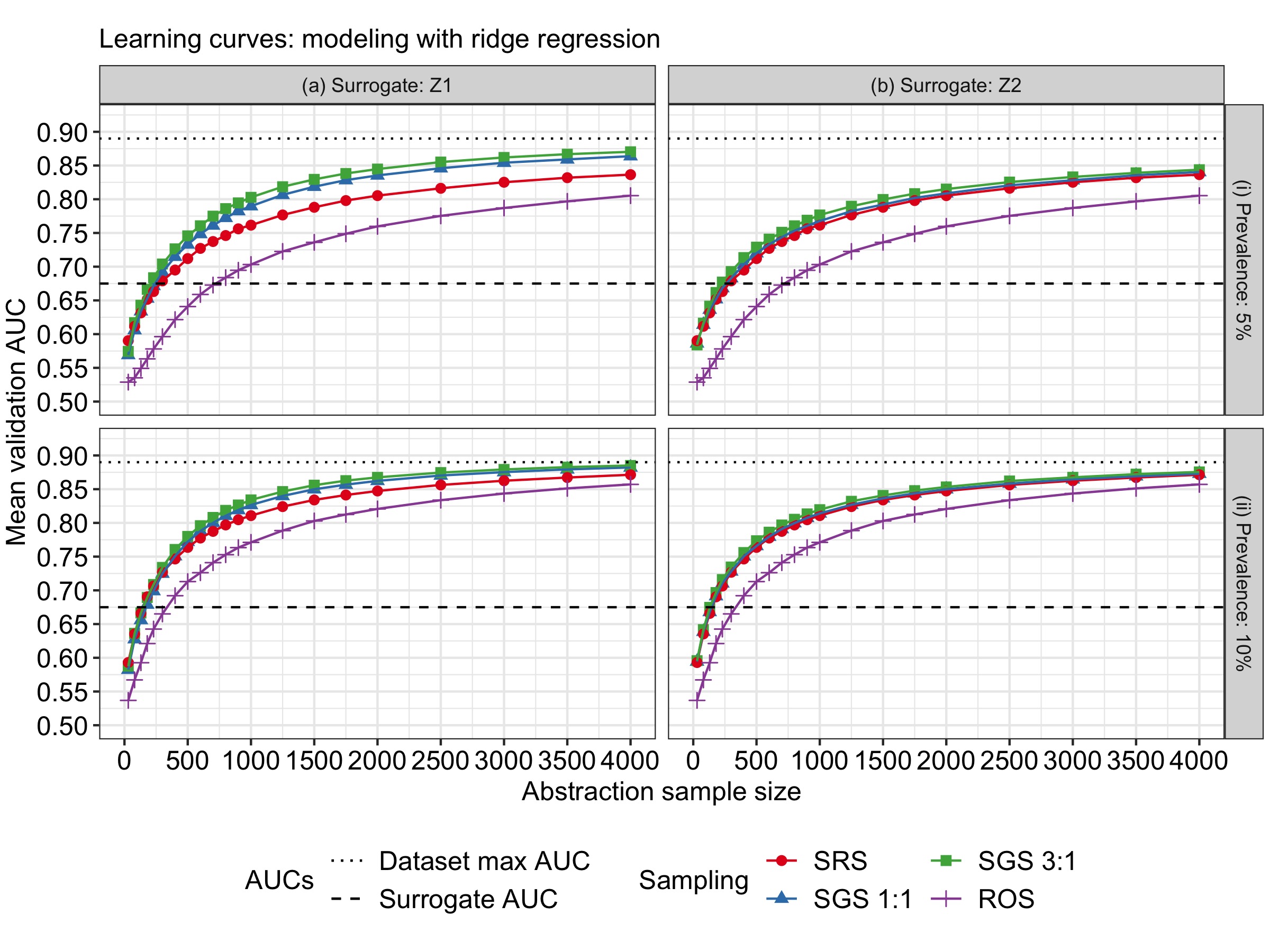}
\label{sgs_fig:cohort_ridge}
\end{figure}

\noindent Figure \ref{sgs_fig:cohort_ridge} illustrates learning curves for modeling with logistic ridge regression, with SGS using surrogate Z1 (Figure \ref{sgs_fig:cohort_ridge}(a)) and Z2 (Figure \ref{sgs_fig:cohort_ridge}(b)). Compared to using lasso regression, the different shapes of learning curves reflected differences in choice of modeling using variable selection versus shrinkage. To achieve a validation AUC of 0.85 for the 5\% outcome prevalence cohort, learning with SRS required an abstraction sample size of at least n=4000 while SGS sampling using surrogate Z1 required about n=2500 (63\% of SRS cost), where using SGS regardless of stratification allocation was a consistent improvement over SRS. On the other hand, SGS sampling using surrogate Z2 had almost the same sample size requirement as with SRS, again emphasizing the importance of surrogate specificity for sampling. Similar conclusions were observed for the 10\% outcome prevalence cohort. When modeling with ridge regression, ROS was consistently worse than SRS without case replication. One possible explanation is that ridge regression reduces the estimation variance of classification through intentionally biased estimates. With over-sampling, while modeling bias increases, variation remains the same as case replication does not provide additional information, therefore resulting in lower generalizable prediction accuracy.

\subsection{Design impact on model validation}

\subsubsection{Set-up}

\noindent For practical machine-learning, often a single sample is collected and split into training/validation. We compare the design impact on the bias and variance in estimating model generalizable accuracy using SRS and SGS sampling. Similar to the set up for model training, but now we focus on specifically on the scenario of prevalence = 5\%, and a lasso regression model trained with $n_{train} = 5000$ samples collected using SGS 1:1 sampling and surrogate Z1 (40\% sensitivity, 95\% specificity for outcome $Y$). We estimated the generalizable AUC of this model using validation samples collected with SRS, SGS 1:1, and SGS 3:1 sampling designs over a range of validation sample sizes compared to that as assessed in the full cohort ($N=100,000$). The empirical estimator was used for SRS samples, while the IPW-corrected empirical estimator was used for SGS samples since the uncorrected estimator is biased.

\subsubsection{Results}

\begin{figure}[!htbp]
\centering
\caption{Mean and variance of validation AUC comparing naive empirical estimator using simple random sampling (SRS) and inverse probability weighted (IPW) estimator using surrogate-guided sampling (SGS) with 1:1 or 3:1 ratio of surrogate positives to negatives. Simulations were based on surrogate sensitivity = 40\%, surrogate specificity = 95\%, outcome prevalence = 5\%, and a training sample size of $n_{train}$ = 5000 fitting a Logistic Lasso regression model.}
\includegraphics[width=5in, height=4in]{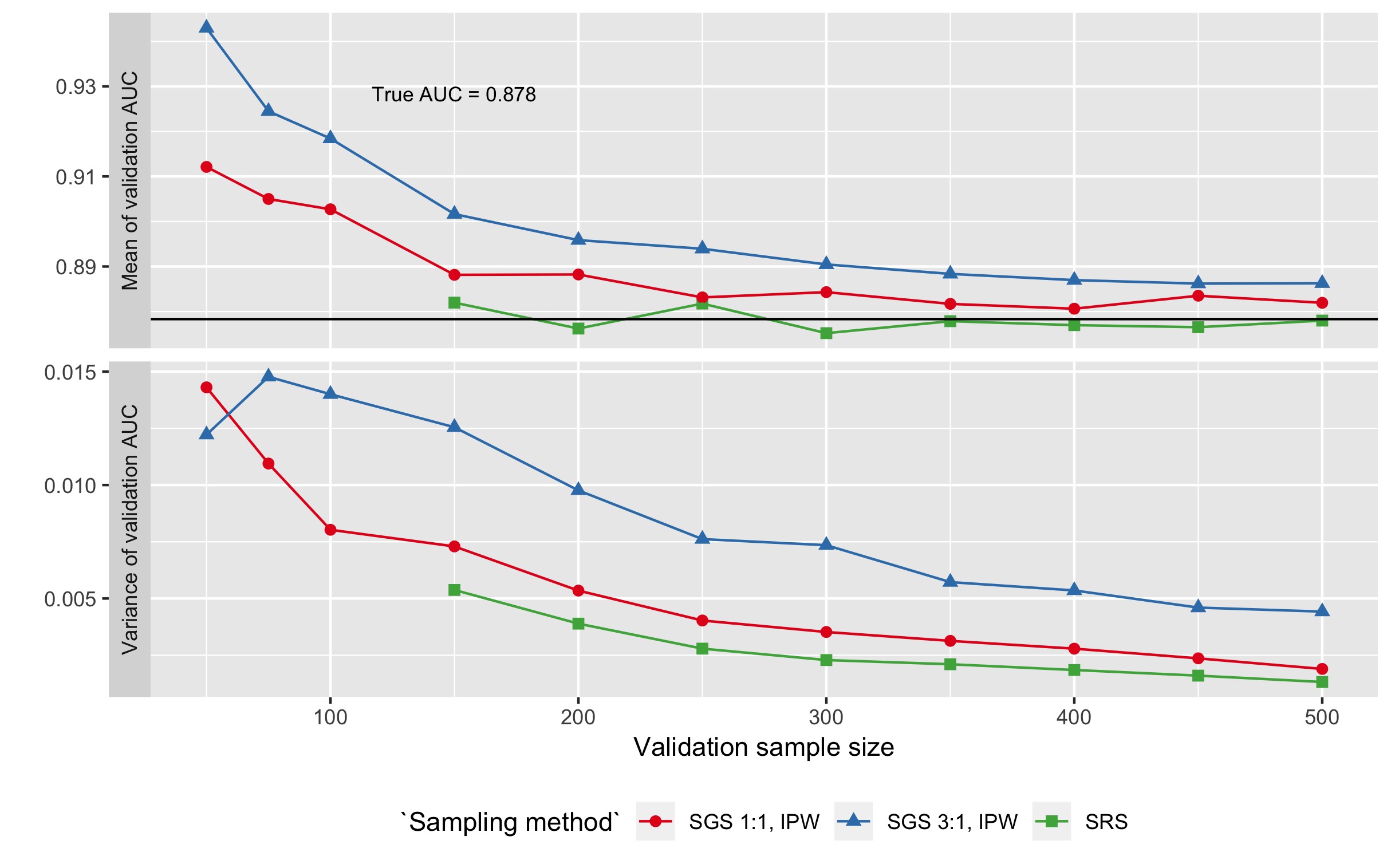}
\label{sgs_fig:sims_validation}
\end{figure}

\noindent Figure \ref{sgs_fig:sims_validation} illustrates the bias and variance of estimating a true AUC of 0.88 using various sampling designs. For small validation sample sizes (here $n<150$), estimating validation AUC was unstable using SRS due to very few cases (e.g. for $n=100$ SRS provides on average 5 true cases), while using SGS estimation of validation AUC was possible although there is noticeable bias. For larger validation sample sizes (here $n>150$), using both the empirical estimator for SRS samples and the IPW-corrected empirical estimator for SGS samples resulted in unbiased estimation of the true AUC. The variances of these unbiased estimators were comparable between the SRS (empirical AUC estimator) and SGS 1:1 (IPW estimator) designs, but the variance of the IPW estimator using the SGS 3:1 design resulted in noticeably higher variance, likely due to extreme weights.

\subsection{Remarks on simulation results}

\noindent Our results suggest three observed patterns associated with using SGS designs for model training and validation. First, for classification of rare outcomes, using SGS designs provides improvement in model training compared to using SRS, and unbiased estimates of model validation performance can be obtained for example using the IPW estimator. Second, using a more specific surrogate results in a sample with higher information and therefore improved learning compared to using a less specific surrogate. Third, in terms of surrogate strata proportions, including more surrogate positives (e.g. using SGS 3:1 versus SGS 1:1) improved learning slightly, but also reduced efficiency in estimating model validation accuracy. For most cases that we investigated, it appears that using a balanced design of SGS 1:1 provides substantial improvement in model training accuracy yet minimal compromise to the efficiency in estimating model validation accuracy.

\section{Application: Fracture identification from radiology reports}
\label{sec:sgs_data}
\subsection{Data set details}

\noindent Vertebral fractures of the spine can lead to spinal deformity, loss of vertebral height, crowding of internal organs, and loss of muscles, resulting in acute back pain and potentially chronic pain. Diagnosis is usually made through radiographic imaging, such as with plain x-ray or magnetic resonance imaging (MRI). In EMR systems a vertebral fracture finding is natively captured in unstructured text form, and for research a definite fracture status variable requires clinical expert abstraction of associated radiology text reports. Therefore, sampling strategies alternative to the usual SRS may reduce the abstraction burden towards accurate and scalable machine-learning classification of vertebral fracture outcomes.\\

\noindent The LIRE study evaluated the effect of radiology report content on subsequent treatment decisions among adult subjects (\cite{jarvik2015lumbar}). Subjects were eligible for the LIRE study if they had a diagnostic imaging test ordered by their Primary Care Physician (PCP), so all subjects in LIRE had at least one radiology report available from the EMR database. The prevalence of vertebral fractures is estimated to be relatively rare: 3-20\% among primary care subjects seeking care for all reasons (\cite{waterloo2012prevalence}) and expected to be similar among subjects from the LIRE study. Using LIRE data as the ``cohort'', we evaluate the benefit of using SGS designs for outcome label abstraction and subsequent classification model development. 

\subsection{Surrogate creation and sampling design application}

\noindent Together with clinicians, we identified a set of $26$ ICD codes that if present, are highly likely to indicate that a subject was diagnosed with a vertebral fracture; details are in Supplementary Material C. For each subject, we counted how many ICD codes were noted in the EMR within 90 days of cohort entry. In the cohort of 178,333 subjects, 171,592 (96\%) did not have any relevant ICD codes, 3,275 (1.83\%) had one code, 1,303 (0.73\%) had two codes, 758 (0.42\%) had three codes, and 1405 (0.79\%) had more than three codes. Since most subjects did not have any relevant ICD codes and among patients who had at least one relevant ICD code a count of one was the most common count, we defined the enrichment surrogate Z as $Z_i = I(\text{count vertebral fracture ICD codes within 90 days for subject } i \geq 1)$, where 3.78\% of the cohort were considered to be ``surrogate positives".\\

\noindent This abstraction task was nested within a larger abstraction set-up for the LIRE study. The radiology reports of each selected subject were abstracted by two independent clinicians for the presence or absence of vertebral fractures. From the available dataset, data ``marts" for model training and model validation were assembled, each having a sample size of n=500. The validation data mart was selected such that it was representative of the underlying cohort (outcome prevalence = 10\%), while the training data mart was selected based on an SGS 1:1 configuration (outcome prevalence = 47\%). Using the validation data mart, we estimated surrogate marginal sensitivity = 27\%  (95\% C.I. 18\% , 36\% ), specificity = 99\%  (95\% C.I. 99\% , 100\% ), AUC = 0.63 (95\% C.I. 0.58, 0.68), $LR+$ = 26 (95\% C.I. 18, 41), $LR-$ = 1.36 (95\% C.I. 1.22, 1.57), and using an SGS design with 1:1 ratio of $Z=1$ and $Z=0$ had an estimated $O_{ratio}$ = 7.13 (95\% C.I. 5.65, 9.04).

\subsection{Modeling and analysis}

\noindent Features were created by processing radiology report text data using the \texttt{quanteda} package in \texttt{R}. Features were BOW unigrams excluding typical English stopwords as well as terms that were very rare ($<5\%$ of all reports) or common ($>90\%$ of all reports). We used the term-frequency inverse-document frequency (TF-IDF) representation for BOW (\cite{salton1988term}), which incorporates information about keyword importance both locally (within a single report) as well as globally (across all reports). For a collection of $N$ reports denoted $d_1, \hdots, d_N$, the set of $p$ terms denoted $T = \{t_1, \hdots, t_p\}$ was obtained from concatenating unique words from all reports. Then the TF-IDF feature matrix $\mathbf{X}$ contains elements $X_{ij} = TF(d_i,t_j) \times IDF(t_j)$, with term frequency TF defined as $TF(t_j, d_i) = 1 + \log(1 + \frac{Count(t_j \in d_i)}{|d_i|})$ and inverse document frequency IDF defined as $IDF(t_j) = \log \left( \frac{N}{\sum\limits_{i=1}^{N} I(t_j \in d_i)} \right)$. In addition to text-features, we also included the binary enrichment surrogate $Z$ as a predictor, for a total of $p=298$ features.\\

\noindent To investigate the design effect on model prediction accuracy, we drew $B=1000$ bootstrap samples of sizes $n=100, 250, 500$ from the training data mart stratified by surrogate status. To simulate the SRS design, we drew samples according to an ``inverse SGS'' design from the training data mart, where surrogate positives were under-included with the sampling probabilities. To simulate the SGS design, we drew samples randomly from the training data mart. For each simulated sample, we fitted Lasso logistic regression selecting regularization parameter based on minimizing the average ten-fold cross-validated error using an AUC loss function. Resulting estimated model parameters were then applied to the validation sample to obtain estimates of the validation AUC. For each sampling design (SRS and SGS) and for each sample size, we reported mean validation AUC and 95\% bootstrap confidence intervals.\\

\noindent Data analysis results are shown in Table \ref{sgs_tb:data_results}, where for the same sample size, using samples drawn with SGS resulted in higher average validation AUC. Such differences was most pronounced for an abstraction sample size of $n=250$, where the AUC of SGS was 0.86 while that of SRS was only 0.74, a difference of 0.12. These results suggest that for applications similar to the LIRE data application example, if only a modest sample size can be collected (e.g. $n=250$), sample collection based on an SGS design is more resource efficient for model building compared to using SRS. Code and a derived dataset to reproduce analyses from this section are available at https://github.com/wlktan/surrogate\_guided\_sampling\_designs.

\section{Discussion}
\label{sec:sgs_discussion}
\noindent Motivated by sampling frameworks from epidemiology and machine-learning, we formalized a design strategy for abstraction selection and label collection of rare outcomes through a two-phase stratified sampling framework. One concern may be whether sampling on a highly specific surrogate could result in a dataset that is sufficiently representative of all possible outcome subgroups. For example, in the vertebral fracture data application, while requiring at least two instead of one ICD codes may further increase surrogate specificity, such a strategy could have resulted in a sample with mostly chronic fractures and not acute fractures. A possible solution may implement a ``tiered" surrogate, using sub-samples defined by variables to balance specificities and case representativeness (e.g. $>2$, $1$, $0$ counts of ICD codes).\\

\noindent In this work we primarily demonstrated design impact on model training and validation on a specific metric, the AUC. However, we expect similar conclusions to other accuracy metrics, following conclusions from the machine-learning literature on the impact of class balance on learning (\cite{weiss2001effect}, \cite{batista2004study}). We comment that by design, since training samples are intentionally enriched with true cases, resulting models may not necessarily be well calibrated. We suggest viewing the SGS framework as a method to train prediction models first to obtain good discrimination, and then conducting post-processing calibration for example by using Platt scaling (\cite{platt1999probabilistic}) or isotonic regression (\cite{zadrozny2002transforming}) of resulting predicted probabilities.\\

\noindent Anchored in the proposed SGS design framework, future work may formally investigate methodological and practical questions related to full study planning such as formal sample size calculations. Once relevant trade-offs are carefully defined, appropriate sample size calculations may then proceed taking into account the need of both model training and model validation. Other future work should include: investigating the appropriateness of the SGS framework for outcomes that are much rarer than what we considered (5\%); characterizing design effects on machine-learning problems beyond binary classification; as well as determining best practices for sampling in the presence of site heterogeneity. Ultimately, our hope is to encourage careful statistical and study design thinking when assembling labeled data sets for machine-learning model training and validation, especially when considering the non-trivial abstraction cost in obtaining such labels.

\section{Supplementary Material}
Mathematical details for Sections \ref{sgs_sec:methods_2} and \ref{sgs_sec:methods_3} are in Supplementary Material A and B respectively; details for data elements used to create surrogates as described in Section \ref{sec:sgs_data} are in Supplementary Material C. The code for the data example in Section \ref{sec:sgs_data}, as well as a derived dataset, are both available at https://github.com/wlktan/surrogate\_guided\_sampling\_designs.

\section*{Acknowledgments}
We acknowledge the following grant support for the conduct of this research:  NIH grant P30 AR072572; and NIH grant UL1 TR002319. The content is solely the responsibility of the authors and does not necessarily represent the official views of the National Institutes of Health. \\
     \\
{\it Conflict of Interest}: None declared.

\singlespacing 
\footnotesize 
\bibliographystyle{biorefs}
\bibliography{master}

\begin{thebibliography}{99}

\bibitem[Agarwal \emph{and others}(2016)Agarwal, Podchiyska, Banda, Goel,
  Leung, Minty, Sweeney, Gyang and Shah]{agarwal2016learning}
\textsc{Agarwal, Vibhu, Podchiyska, Tanya, Banda, Juan~M, Goel, Veena, Leung,
  Tiffany~I, Minty, Evan~P, Sweeney, Timothy~E, Gyang, Elsie and Shah,
  Nigam~H}. (2016).
\newblock Learning statistical models of phenotypes using noisy labeled
  training data.
\newblock {\em Journal of the American Medical Informatics
  Association\/}~\textbf{23}(6), 1166--1173.

\bibitem[Batista \emph{and others}(2004)Batista, Prati and
  Monard]{batista2004study}
\textsc{Batista, Gustavo~EAPA, Prati, Ronaldo~C and Monard, Maria~Carolina}.
  (2004).
\newblock A study of the behavior of several methods for balancing machine
  learning training data.
\newblock {\em ACM Sigkdd Explorations Newsletter\/}~\textbf{6}(1), 20--29.

\bibitem[Breslow and Chatterjee(1999)Breslow and Chatterjee]{breslow1999design}
\textsc{Breslow, Norman~E and Chatterjee, Nilanjan}. (1999).
\newblock Design and analysis of two-phase studies with binary outcome applied
  to wilms tumour prognosis.
\newblock {\em Journal of the Royal Statistical Society: Series C (Applied
  Statistics)\/}~\textbf{48}(4), 457--468.

\bibitem[Carroll \emph{and others}(2012)Carroll, Thompson, Eyler, Mandelin,
  Cai, Zink, Pacheco, Boomershine, Lasko, Xu  et~al.]{carroll2012portability}
\textsc{Carroll, Robert~J, Thompson, Will~K, Eyler, Anne~E, Mandelin, Arthur~M,
  Cai, Tianxi, Zink, Raquel~M, Pacheco, Jennifer~A, Boomershine, Chad~S, Lasko,
  Thomas~A, Xu, Hua  \emph{and others}}. (2012).
\newblock Portability of an algorithm to identify rheumatoid arthritis in
  electronic health records.
\newblock {\em Journal of the American Medical Informatics
  Association\/}~\textbf{19}(e1), e162--e169.

\bibitem[Chapman \emph{and others}(2001)Chapman, Fizman, Chapman and
  Haug]{chapman2001comparison}
\textsc{Chapman, Wendy~Webber, Fizman, Marcelo, Chapman, Brian~E and Haug,
  Peter~J}. (2001).
\newblock A comparison of classification algorithms to automatically identify
  chest x-ray reports that support pneumonia.
\newblock {\em Journal of biomedical informatics\/}~\textbf{34}(1), 4--14.

\bibitem[Chatterjee \emph{and others}(2003)Chatterjee, Chen and
  Breslow]{chatterjee2003pseudoscore}
\textsc{Chatterjee, Nilanjan, Chen, Yi-Hau and Breslow, Norman~E}. (2003).
\newblock A pseudoscore estimator for regression problems with two-phase
  sampling.
\newblock {\em Journal of the American Statistical
  Association\/}~\textbf{98}(461), 158--168.

\bibitem[Chawla \emph{and others}(2002)Chawla, Bowyer, Hall and
  Kegelmeyer]{chawla2002smote}
\textsc{Chawla, Nitesh~V, Bowyer, Kevin~W, Hall, Lawrence~O and Kegelmeyer,
  W~Philip}. (2002).
\newblock Smote: synthetic minority over-sampling technique.
\newblock {\em Journal of artificial intelligence research\/}~\textbf{16},
  321--357.

\bibitem[Choi(1998)Choi]{choi1998slopes}
\textsc{Choi, Bernard~CK}. (1998).
\newblock Slopes of a receiver operating characteristic curve and likelihood
  ratios for a diagnostic test.
\newblock {\em American Journal of Epidemiology\/}~\textbf{148}(11),
  1127--1132.

\bibitem[Esteva \emph{and others}(2017)Esteva, Kuprel, Novoa, Ko, Swetter, Blau
  and Thrun]{esteva2017dermatologist}
\textsc{Esteva, Andre, Kuprel, Brett, Novoa, Roberto~A, Ko, Justin, Swetter,
  Susan~M, Blau, Helen~M and Thrun, Sebastian}. (2017).
\newblock Dermatologist-level classification of skin cancer with deep neural
  networks.
\newblock {\em Nature\/}~\textbf{542}(7639), 115--118.

\bibitem[Gronsbell \emph{and others}(2019)Gronsbell, Minnier, Yu, Liao and
  Cai]{gronsbell2019automated}
\textsc{Gronsbell, Jessica, Minnier, Jessica, Yu, Sheng, Liao, Katherine and
  Cai, Tianxi}. (2019).
\newblock Automated feature selection of predictors in electronic medical
  records data.
\newblock {\em Biometrics\/}~\textbf{75}(1), 268--277.

\bibitem[He and Garcia(2009)He and Garcia]{he2009learning}
\textsc{He, Haibo and Garcia, Edwardo~A}. (2009).
\newblock Learning from imbalanced data.
\newblock {\em IEEE Transactions on knowledge and data
  engineering\/}~\textbf{21}(9), 1263--1284.

\bibitem[Horvitz and Thompson(1952)Horvitz and
  Thompson]{horvitz1952generalization}
\textsc{Horvitz, Daniel~G and Thompson, Donovan~J}. (1952).
\newblock A generalization of sampling without replacement from a finite
  universe.
\newblock {\em Journal of the American statistical
  Association\/}~\textbf{47}(260), 663--685.

\bibitem[Jarvik \emph{and others}(2015)Jarvik, Comstock, James, Avins,
  Bresnahan, Deyo, Luetmer, Friedly, Meier, Cherkin  et~al.]{jarvik2015lumbar}
\textsc{Jarvik, Jeffrey~G, Comstock, Bryan~A, James, Kathryn~T, Avins,
  Andrew~L, Bresnahan, Brian~W, Deyo, Richard~A, Luetmer, Patrick~H, Friedly,
  Janna~L, Meier, Eric~N, Cherkin, Daniel~C  \emph{and others}}. (2015).
\newblock Lumbar imaging with reporting of epidemiology (lire)—protocol for a
  pragmatic cluster randomized trial.
\newblock {\em Contemporary clinical trials\/}~\textbf{45}, 157--163.

\bibitem[King and Zeng(2001)King and Zeng]{king2001logistic}
\textsc{King, Gary and Zeng, Langche}. (2001).
\newblock Logistic regression in rare events data.
\newblock {\em Political analysis\/}~\textbf{9}(2), 137--163.

\bibitem[Le~Cessie and Van~Houwelingen(1992)Le~Cessie and
  Van~Houwelingen]{le1992ridge}
\textsc{Le~Cessie, Saskia and Van~Houwelingen, Johannes~C}. (1992).
\newblock Ridge estimators in logistic regression.
\newblock {\em Applied statistics\/}, 191--201.

\bibitem[Little and Rubin(2014)Little and Rubin]{little2014statistical}
\textsc{Little, Roderick~JA and Rubin, Donald~B}. (2014).
\newblock {\em Statistical analysis with missing data\/}. John Wiley \& Sons.

\bibitem[McIsaac and Cook(2014)McIsaac and Cook]{mcisaac2014response}
\textsc{McIsaac, Michael~A and Cook, Richard~J}. (2014).
\newblock Response-dependent two-phase sampling designs for biomarker studies.
\newblock {\em Canadian Journal of Statistics\/}~\textbf{42}(2), 268--284.

\bibitem[Neyman(1934)Neyman]{neyman1934two}
\textsc{Neyman, Jerzy}. (1934).
\newblock On the two different aspects of the representative method: the method
  of stratified sampling and the method of purposive selection.
\newblock {\em Journal of the Royal Statistical Society\/}~\textbf{97}(4),
  558--625.

\bibitem[Pakhomov \emph{and others}(2005)Pakhomov, Buntrock and
  Chute]{pakhomov2005prospective}
\textsc{Pakhomov, Serguei~V, Buntrock, James and Chute, Christopher~G}. (2005).
\newblock Prospective recruitment of patients with congestive heart failure
  using an ad-hoc binary classifier.
\newblock {\em Journal of biomedical informatics\/}~\textbf{38}(2), 145--153.

\bibitem[Pepe(2003)Pepe]{pepe2003statistical}
\textsc{Pepe, Margaret~Sullivan}. (2003).
\newblock {\em The statistical evaluation of medical tests for classification
  and prediction\/}. Medicine.

\bibitem[Platt \emph{and others}(1999)Platt  et~al.]{platt1999probabilistic}
\textsc{Platt, John  \emph{and others}}. (1999).
\newblock Probabilistic outputs for support vector machines and comparisons to
  regularized likelihood methods.
\newblock {\em Advances in large margin classifiers\/}~\textbf{10}(3), 61--74.

\bibitem[Prentice and Pyke(1979)Prentice and Pyke]{prentice1979logistic}
\textsc{Prentice, Ross~L and Pyke, Ronald}. (1979).
\newblock Logistic disease incidence models and case-control studies.
\newblock {\em Biometrika\/}~\textbf{66}(3), 403--411.

\bibitem[Salton and Buckley(1988)Salton and Buckley]{salton1988term}
\textsc{Salton, Gerard and Buckley, Christopher}. (1988).
\newblock Term-weighting approaches in automatic text retrieval.
\newblock {\em Information processing \& management\/}~\textbf{24}(5),
  513--523.

\bibitem[Sichel(1975)Sichel]{sichel1975distribution}
\textsc{Sichel, Herbert~S}. (1975).
\newblock On a distribution law for word frequencies.
\newblock {\em Journal of the American Statistical
  Association\/}~\textbf{70}(351a), 542--547.

\bibitem[Tibshirani(1996)Tibshirani]{tibshirani1996regression}
\textsc{Tibshirani, Robert}. (1996).
\newblock Regression shrinkage and selection via the lasso.
\newblock {\em Journal of the Royal Statistical Society. Series B
  (Methodological)\/}, 267--288.

\bibitem[Wang \emph{and others}(2009)Wang, Scharfstein, Tan and
  MacKenzie]{wang2009causal}
\textsc{Wang, Weiwei, Scharfstein, Daniel, Tan, Zhiqiang and MacKenzie,
  Ellen~J}. (2009).
\newblock Causal inference in outcome-dependent two-phase sampling designs.
\newblock {\em Journal of the Royal Statistical Society: Series B (Statistical
  Methodology)\/}~\textbf{71}(5), 947--969.

\bibitem[Waterloo \emph{and others}(2012)Waterloo, Ahmed, Center, Eisman,
  Morseth, Nguyen, Nguyen, Sogaard and Emaus]{waterloo2012prevalence}
\textsc{Waterloo, Svanhild, Ahmed, Luai~A, Center, Jacqueline~R, Eisman,
  John~A, Morseth, Bente, Nguyen, Nguyen~D, Nguyen, Tuan, Sogaard, Anne~J and
  Emaus, Nina}. (2012).
\newblock Prevalence of vertebral fractures in women and men in the
  population-based troms{\o} study.
\newblock {\em BMC musculoskeletal disorders\/}~\textbf{13}(1), 3.

\bibitem[Wei and Dunbrack~Jr(2013)Wei and Dunbrack~Jr]{wei2013role}
\textsc{Wei, Qiong and Dunbrack~Jr, Roland~L}. (2013).
\newblock The role of balanced training and testing data sets for binary
  classifiers in bioinformatics.
\newblock {\em PloS one\/}~\textbf{8}(7), e67863.

\bibitem[Weiss and Provost(2001)Weiss and Provost]{weiss2001effect}
\textsc{Weiss, Gary~M and Provost, Foster}. (2001).
\newblock The effect of class distribution on classifier learning: an empirical
  study.

\bibitem[Xue and Hall(2015)Xue and Hall]{xue2015does}
\textsc{Xue, Jing-Hao and Hall, Peter}. (2015).
\newblock Why does rebalancing class-unbalanced data improve auc for linear
  discriminant analysis?
\newblock {\em IEEE transactions on pattern analysis and machine
  intelligence\/}~\textbf{37}(5), 1109--1112.

\bibitem[Yu \emph{and others}(2016)Yu, Chakrabortty, Liao, Cai,
  Ananthakrishnan, Gainer, Churchill, Szolovits, Murphy, Kohane
  et~al.]{yu2016surrogate}
\textsc{Yu, Sheng, Chakrabortty, Abhishek, Liao, Katherine~P, Cai, Tianrun,
  Ananthakrishnan, Ashwin~N, Gainer, Vivian~S, Churchill, Susanne~E, Szolovits,
  Peter, Murphy, Shawn~N, Kohane, Isaac~S  \emph{and others}}. (2016).
\newblock Surrogate-assisted feature extraction for high-throughput
  phenotyping.
\newblock {\em Journal of the American Medical Informatics
  Association\/}~\textbf{24}(e1), e143--e149.

\bibitem[Zadrozny(2004)Zadrozny]{zadrozny2004learning}
\textsc{Zadrozny, Bianca}. (2004).
\newblock Learning and evaluating classifiers under sample selection bias.
\newblock In:  {\em Proceedings of the twenty-first international conference on
  Machine learning\/}. ACM. p. 114.

\bibitem[Zadrozny and Elkan(2002)Zadrozny and Elkan]{zadrozny2002transforming}
\textsc{Zadrozny, Bianca and Elkan, Charles}. (2002).
\newblock Transforming classifier scores into accurate multiclass probability
  estimates.
\newblock In:  {\em Proceedings of the eighth ACM SIGKDD international
  conference on Knowledge discovery and data mining\/}. ACM. pp.\  694--699.

\bibitem[Zhao \emph{and others}(2009)Zhao, Lawless and
  McLeish]{zhao2009likelihood}
\textsc{Zhao, Yang, Lawless, Jerald~F and McLeish, Donald~L}. (2009).
\newblock Likelihood methods for regression models with expensive variables
  missing by design.
\newblock {\em Biometrical Journal: Journal of Mathematical Methods in
  Biosciences\/}~\textbf{51}(1), 123--136.

\bibitem[Zhao \emph{and others}(2012)Zhao, Lawless and McLeish]{zhao2012design}
\textsc{Zhao, Yang, Lawless, Jerald~F and McLeish, Donald~L}. (2012).
\newblock Design and relative efficiency in two-phase studies.
\newblock {\em Journal of Statistical Planning and
  Inference\/}~\textbf{142}(11), 2953--2964.

\end{thebibliography}

\clearpage 

\normalsize 
\clearpage 

\clearpage

\clearpage

\clearpage

\begin{table}[h!]
\centering
\caption{Average validation AUC (95\% C.I.) for various training sample sizes, based on B=1000 bootstrap resamples, for illustration of surrogate-guided sampling (SGS) designs on radiology reports drawn from the LIRE data set.} 
\label{sgs_tb:data_results}
\begin{tabular}{ccc}
Training sample size & $\hat{AUC}(\mathbf{D}^{SRS}(n))$ & $\hat{AUC}(\mathbf{D}^{SGS}(n))$ \\ 
  \hline
100 & 0.68 (0.50, 0.90) & 0.76 (0.63, 0.88) \\ 
  250 & 0.74 (0.50, 0.92) & 0.86 (0.79, 0.91) \\ 
  500 & 0.83 (0.50, 0.92) & 0.88 (0.85, 0.91) \\ 
  \hline
\end{tabular}
\end{table}

\newpage 
\appendix
\section{Proof of Theorem 2.1}
\label{sgs_sec:appendix_1}
\setcounter{equation}{0}

\noindent Denote the cohort data as $\mathcal{D} = (\mathbf{X},Y)$, consisting of features $\mathbf{X}$ (implicitly also including the surrogate $Z$), and binary outcomes $Y$. From $\mathcal{D}$, units (typically subjects) are selected to form training and validation samples.

\subsection{Preliminaries}

\noindent In $\mathcal{D}$, let features $\mathbf{X}$ have mean $\mu_{x|y}$ and covariance $\mathbf{\Sigma}_{x|y}$ conditioned on true outcomes $y \in \{0,1\}$. Assume that $\mathbf{X}$ and $Y$ are related through a logistic regression mean model. To estimate regression coefficients, a sample $\mathbf{D}^{S}(n)$ needs to be drawn from $\mathcal{D}$. Then, based on theory from generalized linear models, the resulting estimate $\hat{\beta}$ has the following first and second moments:

\beq\bal 
E^{\mathbf{D}^{S}(n)}[\hat{\beta}] &= \beta + Bias^{\mathbf{D}^{S}(n)}(\hat{\beta})\\
Var^{\mathbf{D}^{S}(n)}(\hat{\beta}) &= ( {\mathbf{X}^s}^T \mathbf{W} \mathbf{X}^s)^{-1}.
\label{eq:glm_theory}
\eal\eeq 

\noindent In \eqref{eq:glm_theory}, $\mathbf{W} = Diag(p_i(1-p_i))$, where $p_i = P(Y_i=1|\mathbf{X},S_i=1;\beta)$ estimates the average probabilities resulting from the sigmoidal transformation of training sample linear predictions. The terms in \eqref{eq:glm_theory} are accurate up to second order approximations. In estimating the regression parameters, denote the bias $Bias^{\mathbf{D}^{S}(n)}(\hat{\beta})$ as $\bm{B}{(\hat{\beta}^S(n))}$ and variance $Var^{\mathbf{D}^{S}(n)}(\hat{\beta})$ as $\bm{V}{(\hat{\beta}^S(n))}$, then both $\bm{B}{(\hat{\beta}^S(n))}$ and $\bm{V}{(\hat{\beta}^S(n))}$ depend on the training sample $\mathbf{D}^{S}(n)$ through sample size $n$ and sampling design $S$. To evaluate the resulting classification model, we use a large validation sample, obtained using simple random sampling from $\mathcal{D}$. Denote the true linear predictions in the validation sample as $\eta := \mathbf{X}^{v}\beta$, with distribution

\beqa\bal 
\mathbf{X}^{v} \beta &\sim N(\mu_y, \sigma_y^2) \\
\mu_y &= \mu^T_{x|y} \beta ;& \:\:
\sigma_y^2 &= \beta^T \mathbf{\Sigma}_{x|y} \beta
\eal\eeqa 

\noindent for $y \in \{0,1\}$. Under the bi-normal ROC assumption, the AUC is

\beqa\bal 
AUC = \Phi(\sqrt{R_{AUC}}) = \Phi \left( \sqrt{\dfrac{(\mu_1 - \mu_0)^2}{\sigma_1^2 + \sigma_0^2}} \right).
\eal\eeqa

\noindent In the classification setting, coefficients are estimated from the training sample $\mathbf{D}^{S}(n)$, where $\mathbf{D}^{S}(n)$ is generated with sampling design $S$ and with training sample size $n$. We use $AUC(Y|\mathbf{D}^{S}(n))$ to denote an indexing of resulting validation AUC, where

\beq\bal 
AUC(Y|\mathbf{D}^{S}(n)) &= \Phi(\sqrt{R_{AUC}(\mathbf{D}^{S}(n))}) \\
R_{AUC}(\mathbf{D}^{S}(n)) &= \dfrac{(\hat{\mu}_1 - \hat{\mu}_0)^2}{\hat{\sigma}_1^2+ \hat{\sigma}_0^2}. 
\label{sgs_appendix_eq:app2_1}
\eal\eeq 

\noindent In \eqref{sgs_appendix_eq:app2_2}, the notation $\hat{.}$ and $\mathbf{D}^{S}(n)$ indicates that the estimation of $\hat{\beta}$ is from $\mathbf{D}^{S}(n)$. This proof outlines $AUC(\mathbf{D}^{S}(n))$ in terms of training sample composition.

\subsection{Mean and variances of validation sample linear predictions}

\noindent In the large and representative validation sample, for $y \in \{0,1\}$, the mean of the estimated linear predictions is 

\beq\bal 
\hat{\mu}_y &= E^{\mathbf{D}^{S}(n),\mathbf{X}^v}[\mathbf{X}^{v}\hat{\beta}|Y^{v} = y] \\
&= E^{\mathbf{X}^v} E^{\mathbf{D}^{S}(n)|\mathbf{X}^v}[\mathbf{X}^{v}\hat{\beta}|Y^{v} = y] \\
&= E^{\mathbf{X}^v} [\mathbf{X}^{v} (\beta + \bm{B}{(\hat{\beta}^S(n))})|Y^{v} = y] \\
&= \mu^T_{x|y}(\beta + \bm{B}{(\hat{\beta}^S(n))}). \label{sgs_appendix_eq:app2_2}
\eal\eeq 

\noindent where the double expectation is due to the dependence on validation sample features $\mathbf{X}^v$ as well as training sample estimated coefficients $\hat{\beta}$. Similarly, the variance of the estimated linear predictions is 

\beq\bal 
\hat{\sigma}_y^2 &= Var^{\mathbf{D}^{S}(n), \mathbf{X}^v}(\mathbf{X}^{v}\hat{\beta}|Y^{v} = y) \\
&= Var^{\mathbf{X}^v} (E^{\mathbf{D}^{S}(n)|\mathbf{X}^v} [\mathbf{X}^{v} \hat{\beta}|Y^{v} = y])  + E^{\mathbf{X}^v} [Var^{\mathbf{D}^{S}(n)|\mathbf{X}^v} (\mathbf{X}^{v}\hat{\beta}|Y^{v} = y)] \label{sgs_appendix_eq:app2_3}
\eal\eeq  

\noindent The first part of the right hand side of \eqref{sgs_appendix_eq:app2_3} is

\beq\bal 
Var^{\mathbf{X}^v} (E^{\mathbf{D}^{S}(n)|\mathbf{X}^v} [\mathbf{X}^{v}\hat{\beta}|Y^{v} = y])
&= Var^{\mathbf{X}^v} (\mathbf{X}^{v} (\beta + \bm{B}{(\hat{\beta}^S(n))}) |Y^{v} = y) \\
&= (\beta + \bm{B}{(\hat{\beta}^S(n))})^T \mathbf{\Sigma}_{x|y} (\beta + \bm{B}{(\hat{\beta}^S(n))}), \label{sgs_appendix_eq:app2_4}
\eal\eeq 

\noindent and the second part of the right hand side of \eqref{sgs_appendix_eq:app2_3} is

\beq\bal 
E^{\mathbf{X}^v} [ Var^{\mathbf{D}^{S}(n)|\mathbf{X}^v} (X^{v }\hat{\beta}|Y^{v} = y)] 
&= E^{\mathbf{X}^v} [{\mathbf{X}^{v}}^T \bm{V}{(\hat{\beta}^S(n))} \mathbf{X}^{v} | Y^{v} = y] \\ 
&= trace(\bm{V}{(\hat{\beta}^S(n))} \mathbf{\Sigma}_{x|y}) + \mu_{x|y}^T \bm{V}{(\hat{\beta}^S(n))} \mu_{x|y}, \label{sgs_appendix_eq:app2_5}
\eal\eeq 

\noindent where we have used properties of the expectation of a quadratic form: for $\epsilon \sim (\mu, \mathbf{\Sigma}), E[\epsilon^T \Lambda \epsilon] = trace(\Lambda \mathbf{\Sigma}) + \mu^T \Lambda \mu$. Therefore, combining \eqref{sgs_appendix_eq:app2_4} and \eqref{sgs_appendix_eq:app2_5}, the variance of $\eta$ is

\beq\bal 
\hat{\sigma}_y^2 &= Var^{\mathbf{X}^v} (E^{\mathbf{D}^{S}(n)|\mathbf{X}^v} [\mathbf{X}^{v} \hat{\beta}|Y^{v} = y])  + E^{\mathbf{X}^v} [Var^{\mathbf{D}^{S}(n)|\mathbf{X}^v} (\mathbf{X}^{v}\hat{\beta}|Y^{v} = y)] \\
&= (\beta + \bm{B}{(\hat{\beta}^S(n))})^T \mathbf{\Sigma}_{x|y} (\beta + \bm{B}{(\hat{\beta}^S(n))}) + trace(\bm{V}{(\hat{\beta}^S(n))} \mathbf{\Sigma}_{x|y}) + \mu_{x|y}^T \bm{V}{(\hat{\beta}^S(n))} \mu_{x|y}. \label{sgs_appendix_eq:app2_6}
\eal\eeq

\subsection{Classifier validation AUC in terms of estimation variance}

\noindent Now we plug in values for \eqref{sgs_appendix_eq:app2_1}. WLOG assume that $\mu_{x|y0} = 0$ and that $\mathbf{\Sigma}_{x|y=1} = \mathbf{\Sigma}_{x|y=0} = \mathbf{\Sigma}_{x|y}$. Then, the means and variances of validation sample linear predictions among cases (Y=1) and controls (Y=0) are respectively

\beq\bal 
\hat{\mu}_1 &= \mu^T_{x|y1}(\beta + \bm{B}{(\hat{\beta}^S(n))}) \\
\hat{\mu}_0 &= 0 \\
\hat{\sigma}_1^2 &= (\beta + \bm{B}{(\hat{\beta}^S(n))})^T \mathbf{\Sigma}_{x|y}(\beta + \bm{B}{(\hat{\beta}^S(n))}) + 
trace(\mathbf{c}\mathbf{\Sigma}_{x|y}) + \mu_{x|y1}^T \bm{V}{(\hat{\beta}^S(n))} \mu_{x|y1} \\
\hat{\sigma}_0^2 &= (\beta + \bm{B}{(\hat{\beta}^S(n))})^T \mathbf{\Sigma}_{x|y}(\beta + \bm{B}{(\hat{\beta}^S(n))}) + 
trace(\bm{V}{(\hat{\beta}^S(n))}\mathbf{\Sigma}_{x|y}).
\eal\eeq 

\noindent Thus, the numerator in \eqref{sgs_appendix_eq:app2_1} is the square of 

\beq\bal 
\hat{\mu}_1 - \hat{\mu}_0 &= \mu_{x|y1}(\beta + \bm{B}{(\hat{\beta}^S(n))}),
\label{sgs_appendix_eq:app2_7}
\eal\eeq 

\noindent while the denominator in \eqref{sgs_appendix_eq:app2_1} is

\beq\bal 
\hat{\sigma}_1^2 + \hat{\sigma}_0^2 &= 
2\{ (\beta + \bm{B}{(\hat{\beta}^S(n))})^T \mathbf{\Sigma}_{x|y} (\beta + \bm{B}{(\hat{\beta}^S(n))}) + trace(\bm{V}{(\hat{\beta}^S(n))}\mathbf{\Sigma}_{x|y}) \} + 
\mu_{x|y1}^T \bm{V}{(\hat{\beta}^S(n))} \mu_{x|y1}.
\label{sgs_appendix_eq:app2_8}
\eal\eeq 

\noindent Thus, based on \eqref{sgs_appendix_eq:app2_1}, \eqref{sgs_appendix_eq:app2_7}, and \eqref{sgs_appendix_eq:app2_8}, since $\Phi(.)$ and $\sqrt(.)$ are monotone transformations,

\beqa\bal 
AUC(\mathbf{D}^{s}(n)) = \dfrac{(\mu_{x|y1}(\beta + \bm{B}{(\hat{\beta}^S(n))}))^2}{2((\beta+\bm{B}{(\hat{\beta}^S(n))})^T\mathbf{\Sigma}_{x|y}(\beta+\bm{B}{(\hat{\beta}^S(n))}) + trace(\bm{V}{(\hat{\beta}^S(n))}\mathbf{\Sigma}_{x|y})) + 
\mu_{x|y1}^T \bm{V}{(\hat{\beta}^S(n))} \mu_{x|y1}}.
\eal\eeqa 

\noindent When $\bm{B}{(\hat{\beta}^S(n))} \approx 0$, then since $\beta$, $\mathbf{\mu}_{x|y}$ and $\mathbf{\Sigma}_{x|y}$ are assumed to be ``fixed'' quantities in a large validation sample,

\beqa\bal 
AUC(\mathbf{D}^{s}(n)) \propto \dfrac{1}{trace(\bm{V}{(\hat{\beta}^S(n))} \mathbf{\Sigma}_{x|y})) + \mu_{x|y1}^T \bm{V}{(\hat{\beta}^S(n))} \mu_{x|y1}}.
\eal\eeqa

\newpage 
\setcounter{equation}{0}
\section{Details for Section 2.3}
\label{sgs_sec:appendix_2}

\subsection{Derivation of optimal strata proportions for given desired sample outcome prevalence}

\noindent Denote $P(Y=1|S=1)$ as the desired outcome prevalence in an SGS sample, $PPV_Z$, $NPV_Z$ are the positive and negative predictive values of surrogate $Z$ respectively, and $R = P(Z=1|S=1)$ as the proportion of surrogate positives in the sample. Then, for a given desired sample outcome prevalence

\beqa\bal 
P(Y=1|S=1) &= P(Y=1|S=1,Z=1)P(Z=1|S=1) + P(Y=1|S=1,Z=0)P(Z=0|S=0) \\
&= P(Y=1|Z=1)R + P(Y=1|Z=0)(1-R) \\
&= R(P(Y=1|Z=1) - P(Y-1|Z-0)) + P(Y=1|Z=0) \\ 
&= R[PPV_Z - (1 - NPV_Z)] + (1 - NPV_Z) \\
&= R(PPV_Z + NPV_Z - 1) + 1 - NPV_Z
\eal\eeqa 

\noindent where the first equality follows from Bayes rule and the second equality follows from $S \perp Y | Z$. Rearranging yields

\beqa\bal 
R_{opt} &= \dfrac{P(Y=1|S=1) + NPV_Z - 1}{PPV_Z + NPV_Z -1}
\eal\eeqa 

\noindent where we additionally require that $R_{opt} \in [0,1]$.

\subsection{Derivation of Proposition 2.1}

\noindent For $Y \in \{0,1\}$, $E[Y] = P(Y=1)$. Denote subjects where $S = 1$ as those included in $\mathbf{D}^{SGS}(n)$, the SGS sample selected from the cohort only based on values of $Z$. Thus, $S \perp Y |Z$. The expected case odds in samples collected using SGS is

\beqa\bal
Odds(cases|SGS) 
&= \dfrac{E^{\mathbf{D}^{SGS}(n)}[Y|S=1]}{1-E^{\mathbf{D}^{SGS}(n)}[Y|S=1]} 
= \dfrac{P(Y=1|S=1)}{P(Y=0|S=0)}\\
&= \dfrac{P(Y=1|S=1,Z=1)P(Z=1|S=1) + P(Y=1|S=1,Z=0)P(Z=0|S=1)} {P(Y=0|S=1,Z=1)P(Z=1|S=1) + P(Y=0|S=1,Z=0)P(Z=0|S=1)} \\
&= \dfrac{P(Y=1|Z=1)P(Z=1|S=1) + P(Y=1|Z=0)P(Z=0|S=1)} {P(Y=0|Z=1)P(Z=1|S=1) + P(Y=0|Z=0)P(Z=0|S=1)} \\
&= \dfrac{P(Y=1)}{P(Y=0)} \dfrac{R \dfrac{P(Z=1|Y=1)}{P(Z=1)}  + (1-R) \dfrac{P(Z=0|Y=1)}{P(Z=0)}} {R\dfrac{P(Z=1|Y=0)}{P(Z=1)}  + (1-R)\dfrac{P(Z=0|Y=0)}{P(Z=0)}}\\
&= \dfrac{P(Y=1)}{P(Y=0)} \dfrac{R(1-P(Z=1))Z_{sens} + P(Z=1)(1-R)(1-Z_{sens})}{R(1-P(Z=1))(1-Z_{spec}) + P(Z=1)(1-R)(Z_{spec})} \\
&= \dfrac{P(Y=1)}{P(Y=0)} \dfrac{R Z_{sens} + p_Z (1-R-Z_{sens})}{R(1-Z_{spec}) + p_Z (Z_{spec} - R)}
\eal\eeqa

\noindent where 

\beqa\bal
R &= P(Z=1|S=1)\\
p_Z &= P(Z=1) \\
Z_{sens} &= P(Z=1|Y=1)\\
Z_{spec} &= P(Z=0|Y=0).
\eal\eeqa

\noindent The expected case odds in samples collected using SRS is
\beqa\bal
Odds(cases|SRS) 
&= \dfrac{E^{\mathbf{D}^{SRS}(n)}[Y|S=1]}{1-E^{\mathbf{D}^{SRS}(n)}[Y|S=1]}  \\
&= \dfrac{P(Y=1)}{P(Y=0)}.
\eal\eeqa

\noindent Then, the case/control odd ratio of samples obtained with SGS compared to that of SRS is:

\beq\bal
O_{ratio} &= \dfrac{E^{\mathbf{D}^{SGS}(n)}[Y|S=1]}{1-E^{\mathbf{D}^{SGS}(n)}[Y|S=1]}/\dfrac{E^{\mathbf{D}^{SRS}(n)}[Y|S=1]}{1-E^{\mathbf{D}^{SRS}(n)}[Y|S=1]}\\
&= \dfrac{E^{\mathbf{D}^{SGS}(n)}[Y|S=1]}{1-E^{\mathbf{D}^{SGS}(n)}[Y|S=1]}/\dfrac{P(Y=1)}{P(Y=0)}\\
&= \dfrac{RZ_{sens} + p_Z (1-R-Z_{sens})}{R(1-Z_{spec}) + p_Z (Z_{spec} - R)}. \label{sgs_appendix_eq:app1_1}
\eal\eeq 

\noindent Assume that the outcome is rare, so $P(Y = 1) \approx 0$. Then, a linear approximation of \eqref{sgs_appendix_eq:app1_1} is

\beq\bal
O_{ratio} &= \dfrac{RZ_{sens} + p_Z (1-R-Z_{sens})}{R(1-Z_{spec}) + p_Z (Z_{spec} - R)}\\
&= \dfrac{\dfrac{R }{1-P(Y=1|Z=1) } (LR+) + \dfrac{1-R}{P(Y=0|Z=0)} (LR-) } {\dfrac{R}{1-P(Y=1|Z=1)}+ \dfrac{1-R}{P(Z=0|Y=0)}}\\
\approx& (R) (LR+) + (1-R) (LR-) 
\label{sgs_appendix_eq:app1_2}
\eal\eeq 

\noindent where

\beqa\bal
&LR+ =  \dfrac{P(Z=1|Y=1)}{P(Z=1|Y=0)} = \dfrac{Z_{sens}}{1-Z_{spec}} = \dfrac{\dfrac{P(Y=1|Z=1)}{P(Y=0|Z=1)}}{\dfrac{P(Y=1)}{P(Y=0)}}\\
&LR- =  \dfrac{P(Z=0|Y=1)}{P(Z=0|Y=0)} = \dfrac{1-Z_{sens}}{Z_{spec}}  = \dfrac{\dfrac{P(Y=1|Z=0)}{P(Y=0|Z=0)}}{\dfrac{P(Y=1)}{P(Y=0)}} 
\eal\eeqa 

\noindent $LR+$ and $LR-$ are the likelihood ratios of the surrogate $Z$ in predicting the outcome $Y$ among surrogate positives and negatives, respectively.\\

\newpage 
\section{Details of enrichment surrogate for data application}
\label{sgs_sec:appendix_3}

\noindent Table \ref{sgs_tb:fracture_icd_codes} shows details of the set of ICD codes used to construct an enrichment surrogate which is used for collecting reports that are more likely to contain vertebral fracture. The enrichment surrogate was defined as 

\beqa\bal 
Z_i = I((\text{count vertebral fracture ICD codes in Table \ref{sgs_tb:fracture_icd_codes} within 90 days for subject } i) > 1).
\eal\eeqa 

\begin{table}[h!]
\centering
\caption{Set of International Classification of Disease (ICD) codes used to define enrichment surrogate}
\label{sgs_tb:fracture_icd_codes}
\begin{tabular}{c|l}
ICD code &	Long description \\ \hline
806.25 &	Closed fracture of T7-T12 level with unspecified spinal cord injury \\
806.26	& Closed fracture of T7-T12 level with complete lesion of cord \\
806.27	& Closed fracture of T7-T12 level with anterior cord syndrome \\
806.28 &	Closed fracture of T7-T12 level with central cord syndrome\\
806.29	& Closed fracture of T7-T12 level with other specified spinal cord injury\\
806.35&	Open fracture of T7-T12 level with unspecified spinal cord injury\\
806.39	& Open fracture of T7-T12 level with other specified spinal cord injury\\
806.4&	Closed fracture of lumbar spine with spinal cord injury\\
806.5&	Open fracture of lumbar spine with spinal cord injury\\
806.6	&Closed fracture of sacrum and coccyx with unspecified spinal cord injury\\
806.61&	Closed fracture of sacrum and coccyx with complete cauda equina lesion\\
806.62	&Closed fracture of sacrum and coccyx with other cauda equina injury\\
806.69	& Closed fracture of sacrum and coccyx with other spinal cord injury\\
806.8&	Closed fracture of unspecified vertebral column with spinal cord injury\\
806.9&	Open fracture of unspecified vertebral column with spinal cord injury\\
733.13&	Pathologic fracture of vertebrae\\
805.4&	Closed fracture of lumbar vertebra without mention of spinal cord injury\\
805.5&	Open fracture of lumbar vertebra without mention of spinal cord injury\\
805.6&	Closed fracture of sacrum and coccyx without mention of spinal cord injury\\
805.7	&Open fracture of sacrum and coccyx without mention of spinal cord injury\\
805.8&	Closed fracture of unspecified vertebral column without mention of spinal cord injury\\
805.9	&Open fracture of unspecified vertebral column without mention of spinal cord injury\\
809	&Fracture of bones of trunk, closed\\
809.1&	Fracture of bones of trunk, open\\
V54.17	&Aftercare for healing traumatic fracture of vertebrae\\
V54.27&	Aftercare for healing pathologic fracture of vertebrae\\ \hline 
\end{tabular}
\end{table}

\end{document}